\documentclass[letterpaper]{article} % DO NOT CHANGE THIS
\usepackage{aaai2027}  % DO NOT CHANGE THIS
\usepackage[hyphens]{url}  % DO NOT CHANGE THIS
\usepackage{graphicx} % DO NOT CHANGE THIS
\usepackage{natbib}  % DO NOT CHANGE THIS AND DO NOT ADD ANY OPTIONS TO IT
\usepackage{caption} % DO NOT CHANGE THIS AND DO NOT ADD ANY OPTIONS TO IT
\usepackage{algorithm}
\usepackage{algorithmic}

\usepackage{newfloat}
\usepackage{listings}
\DeclareCaptionStyle{ruled}{labelfont=normalfont,labelsep=colon,strut=off} % DO NOT CHANGE THIS
\floatstyle{ruled}
\newfloat{listing}{tb}{lst}{}
\floatname{listing}{Listing}

\usepackage{booktabs}

\usepackage{amsmath,amssymb} 
\usepackage{amsthm}
\usepackage{multirow}
\newtheorem{Def}{Definition}

\title{FSGR: Mitigating Token Frequency Bias for Fair SID-Based Generative Recommendation}
\author {
    Yuchen Zheng \textsuperscript{\rm 1},
    Sihan Xu \textsuperscript{\rm 2},
    Jingwen Yang \textsuperscript{\rm 1},
    Xiangrui Cai
     \textsuperscript{\rm 1}\corresponding,
    Haiwei Zhang \textsuperscript{\rm 1},
     Xiaojie Yuan \textsuperscript{\rm 1}
}
\affiliations {
    \textsuperscript{\rm 1}VCIP, TMCC, DISSec, College of Computer Science, Nankai University\\
    \textsuperscript{\rm 2}VCIP, DISSec, College of Cryptology and Cyber Science, Nankai University\\
    zhengyuchen@dbis.nankai.edu.cn, 
    xusihan@nankai.edu.cn,
    1120250410@mail.nankai.edu.cn,
    caixr@nankai.edu.cn,
    zhhaiwei@nankai.edu.cn,
    yuanxj@nankai.edu.cn
}
\begin{document}

\maketitle

\begin{abstract}
Semantic ID (SID)-based generative recommendation has recently achieved remarkable success. However, existing methods suffer from a previously overlooked fairness issue, which we term \textbf{Token Frequency Bias}, where high-frequency SID tokens are systematically over-predicted while low-frequency SID tokens are under-predicted. This bias originates from the combined effects of imbalanced semantic codebooks during SID construction, and popularity bias together with the maximum likelihood estimation objective during recommendation training, resulting in unfair exposure across item categories. Existing SID methods mainly focus on improving codebook quality and overlook the impact of token frequency imbalance on downstream recommendation fairness, while LLM debiasing methods often yield suboptimal results when directly applied to SID-based recommendation, due to the hierarchical semantics of SID tokens. To address this issue, we propose \textbf{FSGR}, a fairness optimization framework for SID-based generative recommendation. During SID construction, FSGR employs OT-based Assignment Optimization and Dual-Criteria Re-anchor mechanism to form a more balanced SID representation space. During recommendation training, it adopts a two-stage training strategy and introduces Hierarchical Frequency Calibration for layer-specific fairness fine-tuning. Experiments on three public datasets with three backbone models demonstrate that FSGR mitigates token frequency bias and delivers an average Gini fairness improvement of over 20\% while maintaining competitive recommendation accuracy.

\end{abstract}


\section{Introduction} 
Traditional discriminative recommender systems are limited by item cold-start problem and the poor transferability of learned embedding spaces across domains~\cite{rajput2023recommender}. To overcome these limitations, generative recommendation has emerged as a promising paradigm. Among existing approaches, Semantic ID (SID)-based generative recommendation represents each item as a sequence of discrete semantic tokens~\cite{rajput2023recommender, zheng2024adapting}. By replacing conventional item embeddings with semantic token sequences, SID-based methods exhibit stronger semantic generalization and more efficient retrieval, attracting increasing attention from both academia and industry.

\begin{figure}[!thbp]
  \centering
  \includegraphics[width=\linewidth]{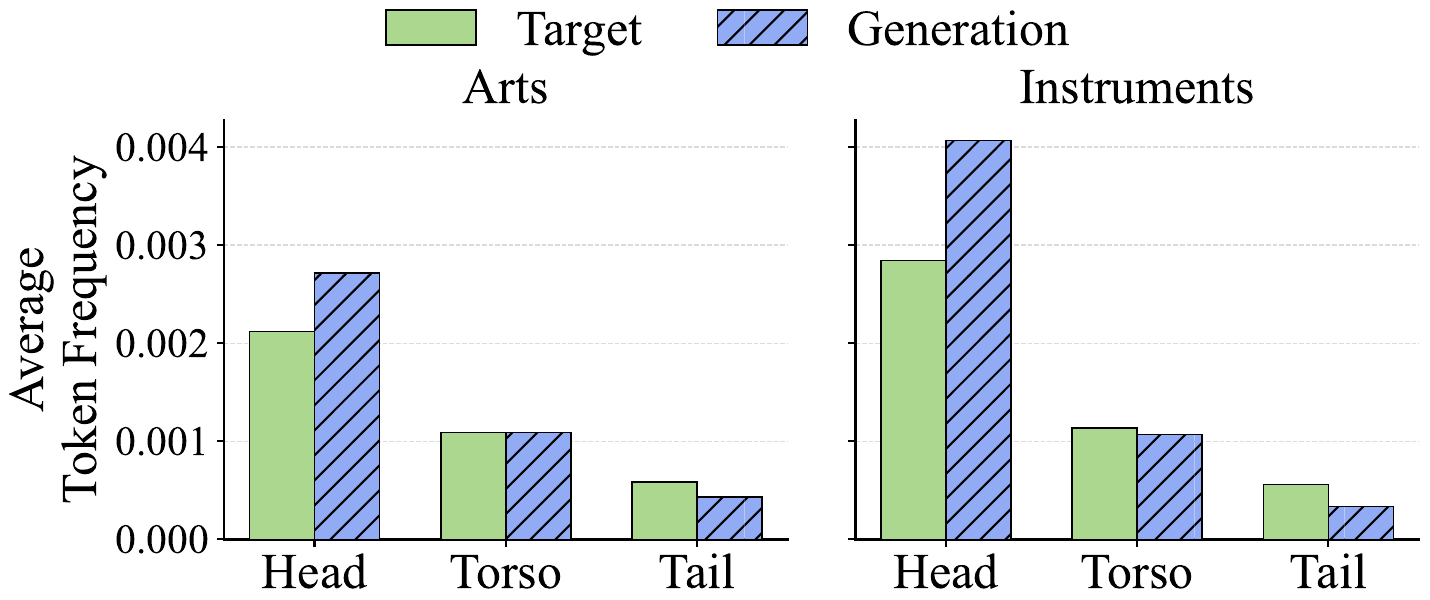}
    \caption{Token group proportions between ground truth and LC-Rec recommendations on two Amazon datasets. Tokens are ranked by training frequency and grouped into Head, Torso, Tail. The results reveal token frequency bias, where high-frequency Head tokens are over-exposed and low-frequency Tail tokens are under-exposed.}
  \label{fig:intro}
\end{figure}

Despite these advantages, SID-based generative recommendation suffers from a critical fairness issue: high-frequency SID tokens are systematically over-predicted, while low-frequency tokens are under-predicted, causing recommendation exposure to concentrate on a small number of frequent token groups and marginalizing long-tail ones. To quantify the practical impact of the SID token imbalance, we conduct a preliminary study. We rank SID tokens according to their frequencies in the training set and divide them into three groups: Head (top 10\%), Torso (middle 50\%), and Tail (bottom 40\%). Using the representative LC-Rec model~\cite{zheng2024adapting}, we generate recommendations on Amazon ``Arts, Crafts and Sewing'' and ``Musical Instruments'' datasets~\cite{ni2019justifying}. We then compare the proportions of the three token groups in the ground-truth labels and the generated recommendations. As illustrated in Figure~\ref{fig:intro}, Head tokens consistently occupy a larger proportion in the recommendation results than in the ground truth, whereas Tail tokens are under-represented. Since Head tokens already dominate the ground-truth distribution, even a modest relative increase results in a significant concentration of exposure on their associated item categories. More importantly, this deviation is systematic rather than random: it consistently favors frequent tokens while suppressing infrequent ones, may create a ``rich-get-richer'' feedback loop that progressively marginalizes long-tail categories~\cite{chang2024cluster}. Inspired by similar observations in natural language processing~\cite{martinez2024mitigating}, we define this phenomenon as \textbf{Token Frequency Bias}.

The emergence of token frequency bias can be attributed to two complementary sources. The first arises during SID construction, where the semantic codebook is inherently imbalanced. Tokens representing broad semantic categories are repeatedly assigned to a large number of items, resulting in higher usage frequencies than other tokens, whereas tokens corresponding to less populated categories are rarely utilized. This imbalance introduces structural bias into the learned semantic space before recommendation training begins. The second arises during recommendation training. On the one hand, popularity bias in the training data causes SIDs of popular items to appear much more frequently than those of long-tail items. On the other hand, the maximum likelihood estimation (MLE) objective amplifies high-frequency training signals~\cite{choi2020, martinez2024mitigating}, assigning higher prediction probabilities to frequent SID tokens while suppressing infrequent tokens. The interaction of these two factors causes the autoregressive generator to favor frequent SID tokens, thus degrading recommendation fairness. However, existing SID studies mainly focus on improving codebook utilization and reducing SID collisions~\cite{kuai2024breaking, hu2026stop}, while overlooking the impact of token frequency imbalance on downstream recommendation fairness. Moreover, dedicated debiasing methods for the LLM stage of SID-based generative recommendation remain unexplored. Although numerous debiasing methods have been proposed for natural language generation and can be applied to SID-based recommendation, they are often suboptimal because they treat all tokens uniformly. In contrast, SID tokens exhibit hierarchical semantics, where different layers encode different levels of semantic granularity and, therefore, require different debiasing strengths.

To address this issue, we propose \textbf{FSGR}, a fairness optimization framework for both SID generation and recommendation training. During SID construction, we introduce an Optimal Transport-based Assignment Optimization (OTA) to encourage balanced token utilization across the codebook within each mini-batch. Furthermore, we design a Dual-Criteria Re-anchor (DCR) mechanism that reuses inactive codewords to refine under-represented and overly crowded regions of the semantic space. Together, OTA and DCR encourage a more balanced SID representation space. During recommendation training, although the generated SIDs are already more balanced, popularity bias in training data and the MLE-based cross-entropy objective still reintroduce frequency bias. To further mitigate this issue, we adopt a two-stage training strategy. In the first stage, the LLM is trained with standard cross-entropy loss to learn semantic mappings. In the second stage, we perform fairness fine-tuning using Hierarchical Frequency Calibration (HFC), which applies different levels of debiasing to different SID layers. The main contributions of this paper are summarized as follows:
\begin{itemize}
\item To the best of our knowledge, this is the first work to identify and define token frequency bias in SID-based generative recommendation, revealing the combination of codebook imbalance, training-data popularity bias, and the intrinsic limitation of the MLE objective causes severe item-side fairness degradation at the SID token level.

\item We propose FSGR, a fairness optimization framework. During SID generation, FSGR balances codebook utilization through OTA and DCR. During recommendation training, FSGR adopts semantic pre-training followed by HFC fine-tuning with layer-specific debiasing to improve SID token-level fairness.

\item Experiments on three public datasets and three backbone models
show that FSGR mitigates token frequency bias, delivering an average Gini fairness improvement of over 20\% while maintaining competitive recommendation accuracy, which verifies its efficacy for fair SID-based generative recommendation.
\end{itemize}

\section{Related Work}

\subsection{Semantic ID-Based Generative Recommendation}
Recently, the paradigm of recommender systems has shifted from traditional discriminative models to generative approaches~\cite{hou2025generative}. To bridge the gap between massive item indices and LLMs, LC-Rec~\cite{zheng2024adapting} employs Residual Quantized Variational Autoencoders (RQ-VAE) to map item features into discrete token sequences known as Semantic IDs (SIDs), and further introduces uniform semantic mapping to mitigate index conflicts in the last SID layer via Sinkhorn-based optimal transport. Subsequent studies further improved the SID construction from different perspectives. ColaRec~\cite{wang2024content} constructs generative item identifiers from collaborative representations and introduces auxiliary indexing and alignment objectives to better integrate the item content with collaborative signals. QuaSID~\cite{hu2026stop} mitigates SID collisions through Hamming-guided repulsion and benign overlap masking, with dual-tower contrastive learning enhancing SID quality. Despite these advances, existing methods mainly focus on improving representation quality and recommendation accuracy, while the fairness issue caused by imbalanced SID token distributions received little attention.

\subsection{Fairness in Recommender Systems}
Fairness has long been an important topic in recommender systems~\cite{wang2023survey, deng2025unbiased}. For example, \citet{zehlike2017fa} adjust recommendation lists to satisfy fairness constraints through re-ranking approaches, \citet{liu2023fairlisa} learn fair item representations through adversarial learning methods, and \citet{chang2024cluster} incorporate fairness objectives into training loss through regularization-based methods.
However, all of these methods operate on recommendation lists, item representations, or ranking objectives, and cannot be directly applied to SID-based generative recommendation, where item exposure is determined by the autoregressive generation of hierarchical SID tokens rather than explicit ranking scores or item embeddings. Thus, mitigating token frequency bias in SID-based generative recommendation remains unexplored. In this work, we focus on item-side fairness, with the goal of mitigating token frequency bias and improving the exposure of items associated with long-tail SID tokens.

\section{Problem Formulation}
Building on the empirical observations in the introduction, we first formally define token frequency bias, then formalize the SID-based generative recommendation task.
\begin{Def}
\textbf{Token Frequency Bias:} In SID-based generative recommendation, the model tends to overestimate the probabilities of high-frequency SID tokens while underestimating those of low-frequency tokens, causing excessive exposure of items associated with frequent tokens and insufficient exposure of items associated with infrequent tokens.
\end{Def}
\noindent This bias is the core fairness problem that we target.

Next, we formalize the core task setting. Let $\mathcal{U}$ and $\mathcal{I}$ denote the sets of users and items, respectively. Each item $v\in\mathcal{I}$ is associated with a feature vector $\mathbf{e}_v\in\mathbb{R}^D$. A Residual Quantized Variational Autoencoder (RQ-VAE) quantizes $\mathbf{e}_v$ into a Semantic ID (SID) sequence $\mathbf{s}_v=[s_v^1,\ldots,s_v^L]$, where $L$ is the number of quantization layers and $s_v^l$ is the codebook token at layer $l$. Given a user's interaction history $\mathcal{H}_u=[v_1,\ldots,v_n]$, SID-based generative recommendation trains a model parameterized by $\Theta$ to autoregressively predict the SID sequence of the target item $v_{n+1}$ by maximizing:
\begin{align}
    \max_{\Theta}\sum_{u\in\mathcal{U}}\sum_{l=1}^{L}
    \log P\!\left(s_{v_{n+1}}^l \mid \mathcal{H}_u, s_{v_{n+1}}^{<l};\Theta\right).
\end{align}

This paper addresses the token frequency bias problem in SID-based LLM generative recommendation by jointly optimizing SID quantization and recommendation generation to reduce token distribution imbalance while preserving recommendation accuracy.

\begin{figure*}[!t]
  \centering
  \includegraphics[width=0.98\linewidth]{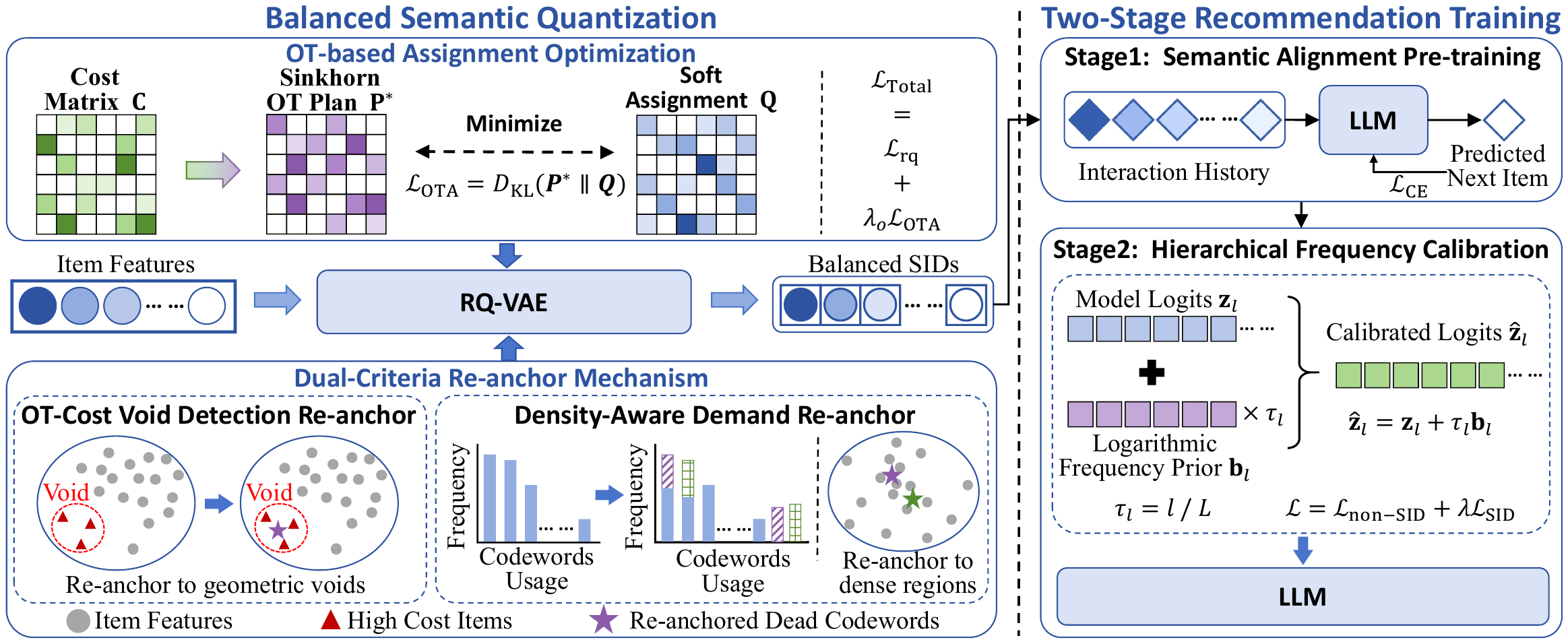}
    \caption{Overview of FSGR. The framework consists of two modules: (1) Balanced Semantic Quantization, which constructs more balanced Semantic IDs via OT-based Assignment Optimization and Dual-Criteria Re-anchor, and (2) Two-Stage Recommendation Training, which performs semantic alignment pre-training followed by Hierarchical Frequency Calibration to mitigate token frequency bias.}
  \label{fig:main}
\end{figure*}
\section{Method}
To mitigate token frequency bias in SID-based generative recommendation, we propose FSGR, a two-component fairness optimization framework. As shown in Figure~\ref{fig:main}, first, OT-based Assignment Optimization (OTA) and Dual-Criteria Re-anchor (DCR) improve codebook utilization during quantization to encourage a more balanced semantic representation space. Building upon this, we adopt a two-stage training strategy: standard cross-entropy first establishes semantic alignment, followed by Hierarchical Frequency Calibration (HFC) that mitigates autoregressive prediction bias via layer-wise granularity calibration while preserving alignment.

\subsection{Balanced Semantic Quantization}
In the conventional way, given an item feature vector $\mathbf{e}_v \in \mathbb{R}^D$, an $L$-layer RQ-VAE method maps it into a discrete token sequence through multi-level residual quantization. At layer $l$, the encoder outputs a residual vector $\mathbf{r}_l$, and the quantizer finds the nearest codeword $s^l$ from the codebook $\mathcal{\tilde{C}}_l \in \mathbb{R}^{K_l \times D}$: $s^l = \arg \min_{k} \|\mathbf{r}_l - \mathbf{e}_k^{l}\|_2^2$, where $\mathbf{e}_k^{l}$ is the embedding of $s^l$, $K_l$ is the vocabulary size of the $l$-th SID layer. The quantized representation for this layer is $\mathbf{q}_l = \mathbf{e}_{k}^{l}$, and the residual passed to the next layer is $\mathbf{r}_{l+1} = \mathbf{r}_l - \mathbf{q}_l$. 
The standard RQ-VAE objective is $\mathcal{L}_{\mathrm{rq}}$ consists of reconstruction loss and commitment loss:
\begin{small}
\begin{align}
\mathcal{L}_{\mathrm{rq}}\!=\!\|\mathbf{e}_v - \hat{\mathbf{e}}_v\|_2^2 
    + \sum_{l=1}^{L} \|\text{sg}[\mathbf{r}_l] - \mathbf{q}_l\|_2^2
    + \beta \sum_{l=1}^{L} \|\mathbf{r}_l - \text{sg}[\mathbf{q}_l]\|_2^2,
\end{align}
\end{small}
where $\hat{\mathbf{e}}_v$ is the reconstructed vector and $\text{sg}[\cdot]$ denotes the stop-gradient operation. However, in conventional RQ-VAE, the utilization of the codebook is typically highly imbalanced. To alleviate this issue, we propose Balanced Semantic Quantization (BSQ), which encourages balanced codebook utilization through OT-based Assignment Optimization and Dual-Criteria Re-anchor mechanism.

\subsubsection{OT-based Assignment Optimization}
In order to address imbalanced codebook utilization, we model the quantization process within each mini-batch as an Optimal Transport (OT) problem, aiming to minimize transport cost while promoting uniform marginal distributions for codebook utilization. 
Formally, given a batch of data $\mathcal{X} \in \mathbb{R}^{B \times D}$ and the codebook $\mathcal{\tilde{C}} \in \mathbb{R}^{K \times D}$, we construct the cost matrix $\mathbf{C}$, where $C_{ik}$ represents the distance between sample $i \in \mathcal{X}$ and the codeword $k \in  \mathcal{\tilde{C}}$. Then we set the sample marginal $\boldsymbol{\mu}$ and codebook target marginal $\boldsymbol{\nu}$ to be uniform: $\boldsymbol{\mu} = \frac{1}{B}\mathbf{1}_B$ and $\boldsymbol{\nu} = \frac{1}{K}\mathbf{1}_K$, and employ the Sinkhorn-Knopp algorithm~\cite{cuturi2013sinkhorn} to solve the entropy-regularized optimal transport plan $\mathbf{P}^* \in \mathbb{R}^{B \times K}$:
\begin{align}
    \mathbf{P}^* = \arg \min_{\mathbf{P} \in \Pi(\boldsymbol{\mu},\boldsymbol{\nu})} \langle \mathbf{P}, \mathbf{C} \rangle - \epsilon H(\mathbf{P}),
\end{align}
where $\Pi(\boldsymbol{\mu},\boldsymbol{\nu})$ is the set of transport matrices satisfying the marginal constraints, $H(\mathbf{P})$ is the entropy regularizer, $\mathbf{P}^*_{ik}$ indicates the optimal probability mass that sample $i$ should assign to codeword $k$.

However, directly applying the discrete transport plan is non-differentiable, thus we define the model’s actual soft assignment matrix $\mathbf{Q}$ as:
\begin{align}
    \mathbf{Q}_{ik} = \frac{\exp(-C_{ik}/\tau_q)}{\sum_{j=1}^{K} \exp(-C_{ij}/\tau_q)},
\end{align}
where $\tau_q$ is the temperature coefficient controlling the smoothness of the soft assignment distribution. We minimize the divergence between the actual soft assignment $\mathbf{Q}$ and the optimal transport plan $\mathbf{P}^*$ through KL divergence:
\begin{align}
    \mathcal{L}_{\mathrm{OTA}} = D_{\mathrm{KL}}(\mathbf{P}^* \parallel \mathbf{Q}) = \sum_{i=1}^{B} \sum_{k=1}^{K} P^*_{ik} \log \frac{P^*_{ik}}{Q_{ik}}.
\end{align}
The total loss function is:
\begin{align}
    \mathcal{L}_{\mathrm{total}} = \mathcal{L}_{\mathrm{rq}} + \lambda_{o} \mathcal{L}_{\mathrm{OTA}},
\end{align}
where $\lambda_{o}$ is a balancing coefficient controlling the strength of the OT regularization, which is gradually increased during training according to a warm-up schedule.

In this step, $\mathcal{L}_{\mathrm{OTA}}$ encourages the assignment distribution to follow a balanced transport plan, thereby improving overall codebook utilization.

\subsubsection{Dual-Criteria Re-anchor Mechanism}

Although OT-based Assignment Optimization improves overall codebook utilization, the learned semantic space may still contain under-represented and overly crowded regions. Meanwhile, some codewords remain inactive for extended periods and receive little or no gradient updates, making them difficult to recover through standard optimization. To address both issues, we propose Dual-Criteria Re-anchor (DCR) mechanism, which periodically identifies dead codewords and re-initializes them to either under-covered or high-density regions according to two complementary criteria. By reusing inactive codewords, DCR not only refines the geometry of the semantic space but also revives dead codewords and improves overall codebook utilization.

Specifically, let the codebook utilization count be $n_k = \sum_{i=1}^{|\mathcal{I}|} \mathbb{I}[s_i = k]$. A codeword $k_d$ is deemed dead if $n_{k_d} < \delta$, where $\delta$ is a threshold and is set to $1$ in this paper. DCR partitions the dead codes and re-anchors them via two strategies:

\paragraph{Strategy A: OT-Cost Void Detection Re-anchor.}To fill geometric voids in the latent feature space, we define the transport cost of training sample $i$ as $\mathrm{Cost}_i = \sum_{k=1}^{K} P^*_{ik} C_{ik}$. A high transport cost indicates that the sample is distant from existing codewords, residing in an under-represented region with geometric voids. We select indices $\mathcal{I}_{\mathrm{void}}$ from samples with the highest costs and re-anchor a portion of dead codes to these samples’ encoded vectors: $\mathbf{e}_{k_d} \leftarrow \mathbf{e}_i, i \in \mathcal{I}_{\mathrm{void}}$.

\paragraph{Strategy B: Density Aware Demand Re-anchor.}
While Strategy A targets under-represented regions, Strategy B focuses on overcrowded ones. We construct a sampling distribution based on codeword utilization frequency and select representative instances from densely populated semantic regions, denoted by $\mathcal{I}_{\mathrm{demand}}$. The remaining dead codewords are re-anchored to the encoded representations of the selected instances: $\mathbf{e}_{k_d} \leftarrow \mathbf{e}_i, i\in\mathcal{I}_{\mathrm{demand}}$, thereby introducing additional codewords into densely populated semantic regions and improving their representation capacity.

Combined with the codebook-level OTA, the codeword-level DCR further refines the codebook geometry through individual codeword re-anchoring, improving the representation of both under-represented and densely populated semantic regions.

\subsection{Two-Stage Recommendation Training}

Although BSQ balances SID representations, token frequency bias persists in recommendation training: popular items dominate interactions, making their SID tokens far more frequent than long-tail ones. And the MLE objective further amplifies this imbalance, biasing the autoregressive decoder toward high-frequency tokens. A straightforward solution is to introduce frequency-aware regularization throughout training. However, doing so interferes with learning semantic correspondence between user behaviors and SIDs, leading to unstable optimization and degraded accuracy. We thus decouple training into two stages: the first learns semantic alignment via standard cross-entropy, while the second conducts fairness fine-tuning with Hierarchical Frequency Calibration to calibrate SID predictions while preserving learned semantic knowledge.

\subsubsection{Semantic Alignment Pre-training}
In the first stage, the model is optimized via standard cross-entropy:
\begin{align} 
\mathcal{L}_{\mathrm{CE}}=-\sum_{l=1}^{L}\log P\left( s^{l} \mid \mathcal{H}, s^{<l} \right),
\end{align} 
where $\mathcal{H}$ is the user's interaction history, $s^{l}$ is the predicted SID at layer $l$. This stage focuses solely on learning semantic correspondence between user behaviors and SID sequences without introducing any frequency-aware constraints.

\subsubsection{Hierarchical Frequency Calibration}

After the model converges to a stable semantic space, we perform fairness-aware fine-tuning through Hierarchical Frequency Calibration (HFC). Let $\mathbf{z}_{l}\in\mathbb{R}^{K_l}$ denote the prediction logits of the $l$-th SID layer. Based on the empirical token distribution of the training set, we compute the logarithmic frequency prior $\mathbf{b}_{l}=-\log(\mathbf{f}_{l}+\epsilon)$, where $\mathbf{f}_{l}$ is the normalized occurrence frequency of SID tokens, $\epsilon$ is a small constant for numerical stability. The calibrated logits are computed as
\begin{align}
\hat{\mathbf{z}}_{l}=\mathbf{z}_{l}+\tau_l\mathbf{b}_{l},
\end{align}
where $\tau_l$ is the layer-specific calibration temperature. By incorporating the empirical frequency prior, HFC reduces the model's reliance on the skewed token distribution induced by the training data.

\begin{table*}[!t]
\small
\setlength{\tabcolsep}{0.95mm}
\begin{tabular}{clccccccccccccccc}
\toprule

& &
\multicolumn{5}{c}{Beauty} &
\multicolumn{5}{c}{Industrial} &
\multicolumn{5}{c}{Software} \\

\cmidrule(lr){3-7}
\cmidrule(lr){8-12}
\cmidrule(lr){13-17}

& Method
& R@5 & R@10 & N@5 & N@10 & G@10
& R@5 & R@10 & N@5 & N@10 & G@10
& R@5 & R@10 & N@5 & N@10 & G@10 \\

\midrule

\multirow{4}{*}{\rotatebox[origin=c]{90}{Tiger}}
& RQ-VAE
& \textbf{0.2673} & 0.2972 & \textbf{0.2365} & \textbf{0.2462} & 0.7310
& \textbf{0.0591} & \textbf{0.0684} & \textbf{0.0509} & \textbf{0.0539} & 0.7565
& \underline{0.1783} & \underline{0.2233} & \underline{0.1561} & \underline{0.1705} & 0.8671 \\

& RT
& 0.2573 & 0.2877 & 0.2276 & 0.2374 & 0.7635
& 0.0553 & \underline{0.0670} & 0.0486 & 0.0524 & \underline{0.6887}
& \textbf{0.1898} & \textbf{0.2260} & \textbf{0.1646} & \textbf{0.1763} & 0.8906 \\

& QuaSID
& 0.2631 & \textbf{0.2980} & 0.2280 & 0.2393 & \underline{0.7043}
& 0.0545 & 0.0644 & 0.0471 & 0.0504 & 0.7111
& 0.1558 & 0.1777 & 0.1392 & 0.1463 & \underline{0.7389} \\

& Our-BSQ 
& \underline{0.2662} & \underline{0.2975} & \underline{0.2311} & \underline{0.2412} & \textbf{0.5494}
& \underline{0.0554} & 0.0653 & \underline{0.0497} & \underline{0.0529} & \textbf{0.5856}
& 0.1722 & 0.2112 & 0.1486 & 0.1613 & \textbf{0.6981} \\

\midrule

\multirow{6}{*}{\rotatebox[origin=c]{90}{Llama3.1-8B}}
& RQ-VAE
& \underline{0.3116} & \underline{0.3234} & \underline{0.2855} & 0.2894 & 0.7174
& 0.0904 & \underline{0.0970} & \underline{0.0839} & \underline{0.0861} & 0.6059
& 0.2139 & 0.2386 & 0.1834 & 0.1913 & 0.8610 \\

& RT
& 0.3035 & 0.3145 & 0.2825 & 0.2862 & 0.7573
& 0.0886 & 0.0947 & 0.0814 & 0.0834 & 0.6022
& \underline{0.2183} & \textbf{0.2551} & \underline{0.1871} & \underline{0.1989} & 0.8570 \\

& QuaSID
& 0.2906 & 0.2964 & 0.2688 & 0.2708 & \underline{0.7122}
& 0.0884 & 0.0938 & 0.0811 & 0.0829 & 0.6818
& 0.1980 & 0.2145 & 0.1759 & 0.1813 & \underline{0.7484} \\

& MiLe
& \textbf{0.3119} & 0.3232 & \textbf{0.2859} & \underline{0.2896} & 0.7281
& \textbf{0.0953} & \textbf{0.1022} & \textbf{0.0880} & \textbf{0.0902} & \underline{0.5931}
& \textbf{0.2194} & \underline{0.2447} & \textbf{0.1932} & \textbf{0.2013} & 0.8609 \\

& WAKL
& 0.3072 & \textbf{0.3276} & 0.2836 & \textbf{0.2903} & 0.7261
& 0.0885 & 0.0952 & 0.0817 & 0.0839 & 0.6029
& 0.2057 & 0.2298 & 0.1816 & 0.1895 & 0.8586 \\

& Our
& 0.3066 & 0.3164 & 0.2801 & 0.2832 & \textbf{0.4976}
& \underline{0.0918} & 0.0967 & 0.0835 & 0.0851 & \textbf{0.3439}
& 0.2145 & 0.2414 & 0.1866 & 0.1952 & \textbf{0.6754} \\

\midrule

\multirow{6}{*}{\rotatebox[origin=c]{90}{Qwen3-8B}}
& RQ-VAE
& 0.3158 & 0.3353 & 0.2895 & 0.2957 & 0.7368
& 0.0908 & 0.0989 & \textbf{0.0839} & \textbf{0.0865} & 0.6559
& 0.2090 & 0.2414 & 0.1840 & 0.1944 & 0.8646 \\

& RT
& 0.3182 & 0.3337 & 0.2929 & 0.2979 & 0.7588
& 0.0889 & 0.0966 & 0.0814 & 0.0839 & 0.6431
& \underline{0.2117} & 0.2463 & 0.1849 & 0.1961 & 0.8568 \\

& QuaSID
& 0.3177 & 0.3305 & 0.2879 & 0.2921 & \underline{0.7154}
& 0.0858 & 0.0926 & 0.0769 & 0.0791 & 0.7054
& 0.2002 & 0.2408 & 0.1749 & 0.1881 & \underline{0.7638} \\

& MiLe
& \underline{0.3195} & \underline{0.3397} & \underline{0.2932} & \underline{0.2998} & 0.7292
& \textbf{0.0920} & \underline{0.0992} & \textbf{0.0839} & 0.0862 & \underline{0.6135}
& 0.2112 & \underline{0.2496} & \textbf{0.1887} & \underline{0.2007} & 0.8602 \\

& WAKL
& 0.3122 & 0.3295 & 0.2848 & 0.2904 & 0.7364
& \underline{0.0919} & \textbf{0.0995} & \textbf{0.0839} & \underline{0.0863} & 0.6371
& 0.2068 & 0.2457 & 0.1797 & 0.1923 & 0.8612 \\

& Our
& \textbf{0.3329} & \textbf{0.3521} & \textbf{0.2960} & \textbf{0.3023} & \textbf{0.5128}
& 0.0891 & 0.0971 & 0.0807 & 0.0833 & \textbf{0.4203}
& \textbf{0.2189} & \textbf{0.2639} & \underline{0.1872} & \textbf{0.2019} & \textbf{0.7127} \\

\bottomrule
\end{tabular}
\caption{Performance comparison with SID generation and LLM token debiasing baselines on Tiger, Llama3.1-8B and Qwen3-8B. Best and second-best results are highlighted in bold and underlined, respectively.}
\label{tab:main}
\end{table*}

Moreover, SID possesses a hierarchical semantic structure: lower layers encode coarse semantic categories, whereas higher layers capture increasingly fine-grained item semantics~\cite{rajput2023recommender}. Applying the same calibration strength to all layers either distorts coarse semantic representations or insufficiently debiases fine-grained predictions. Thus, HFC adopts a layer-aware temperature schedule. The calibration temperature increases progressively across SID layers according to $\tau_l = l / L, ~ l\in\{1,\ldots,L\}$. So that deeper layers receive stronger frequency calibration, while lower layers remain relatively stable. This design aligns with the hierarchical semantics of SID, mitigating token frequency bias while preserving recommendation accuracy.

The calibrated logits are used only for SID prediction positions during the second-stage fine-tuning, while non-SID tokens are optimized using the original logits. The overall objective is
$
\mathcal{L}=\mathcal{L}_{\mathrm{non\text{-}SID}}+\lambda_h\mathcal{L}_{\mathrm{SID}},
$
where
\begin{align}
\mathcal{L}_{\mathrm{non\text{-}SID}} = -\sum_{t\in\Omega_N}\log P(y_t\mid\mathbf{x}), \\
\mathcal{L}_{\mathrm{SID}} = - \sum_{t\in\Omega_S} \log \frac{\exp(\hat z^{(l_t)}_{t,y_t})}{\sum_{j=1}^{K_{l_t}} \exp(\hat z^{(l_t)}_{t,j})}.
\end{align}
Here, $\Omega_N$ and $\Omega_S$ denote prediction positions of non-SID and SID tokens, $\mathbf{x}$ denotes input sequence, $y_t$ is the target token at prediction position $t$, $\hat z$ is the calibrated logit, and $\lambda_h$ balances recommendation learning and frequency calibration.

\section{Experiments}

\subsection{Experimental Setup}
\subsubsection{Dataset}
We evaluated the proposed approach on three subsets of Amazon review data~\cite{ni2019justifying}, including “Luxury Beauty”, “Industrial and Scientific”, “Software”. Each item in these datasets is associated with a title and a description. Following prior work~\cite{zheng2024adapting}, we eliminated users and items with fewer than five interactions. User behavior sequences were constructed in chronological order, with a unified maximum length of 20.

\subsubsection{Baselines and Backbones}
We compared FSGR with three representative SID generation methods: vanilla \textbf{RQ-VAE}~\cite{rajput2023recommender}, \textbf{RT}~\cite{fifty2025restructuring}, and \textbf{QuaSID}~\cite{hu2026stop}, as well as two LLM token debiasing methods, \textbf{MiLe}~\cite{su2024mile} and \textbf{WAKL}~\cite{shrestha2025llm}. Experiments were conducted on three representative generative recommendation backbones: \textbf{TIGER}~\cite{rajput2023recommender}, \textbf{Llama3.1-8B}~\cite{grattafiori2024llama3herdmodels}, and \textbf{Qwen3-8B}~\cite{qwen3technicalreport}.

\subsubsection{Evaluation Metrics} 
We evaluated our method in terms of recommendation performance and fairness. Following previous work, recommendation performance were measured by \textbf{Recall (R@K)} and \textbf{NDCG (N@K)}, while fairness were evaluated using the \textbf{Gini coefficient (G@K)}~\cite{gini1936measure} computed on the generated SID token frequency distribution. A lower Gini coefficient indicates more balanced token exposure and weaker token frequency bias.

\subsubsection{Implementation Details}
We used a 4-layer RQ-VAE with a codebook size of 256 for each layer. Our method was implemented using PyTorch and experiments were conducted on an NVIDIA RTX A6000. The LLM models were finetuned with Low-Rank Adaptation (LoRA)~\cite{hu2022lora}. We optimized the model using AdamW-8bit~\cite{dettmers20218} optimizer and the detailed hyper-parameter settings are provided in the supplementary material and code.

\subsection{Performance Comparison}
Table~\ref{tab:main} reports the overall performance of FSGR. Since TIGER is a Transformer-based recommender rather than an LLM, HFC and LLM token debiasing baselines were evaluated only on Llama and Qwen, whereas TIGER was used only to assess SID generation component. For Llama and Qwen, we compared the complete FSGR framework with both SID generation and LLM token debiasing baselines. On the TIGER backbone, our proposed BSQ yields the lowest Gini scores across all datasets, surpassing other SID generation methods in recommendation fairness. Meanwhile, it retains competitive Recall and NDCG, indicating that balanced codebook utilization effectively mitigates token frequency bias with only marginal impact on recommendation accuracy. For the two LLM backbones (Llama3.1 and Qwen3), the full FSGR framework achieves the best fairness performance, delivering lower Gini coefficients than both SID generation and LLM token debiasing baselines while preserving competitive recommendation accuracy. These results demonstrate that the combination of balanced SID construction and hierarchical frequency calibration alleviates token frequency bias without compromising recommendation performance.

\subsection{Ablation Analysis}

To evaluate the contribution of each component, we conducted an ablation study on Llama and Qwen. We compared four variants: \textbf{w/o BSQ}, which removes Balanced Semantic Quantization, \textbf{w/o HFC}, which removes Hierarchical Frequency Calibration, \textbf{Raw}, which uses neither BSQ nor HFC, and the complete FSGR framework (\textbf{Full}). As shown in Table~\ref{tab:ablation}, the complete framework achieves the lowest Gini scores across all datasets and backbone models, demonstrating that BSQ and HFC complement each other in mitigating token frequency bias. Removing either module leads to a degradation in fairness, confirming that both balanced SID construction and frequency-aware recommendation training are necessary. From the perspective of individual components, removing BSQ generally results in higher Gini scores than the full model, indicating that constructing a balanced semantic codebook provides a stronger foundation for downstream recommendation fairness. Removing HFC also causes an increase in Gini, suggesting that frequency bias introduced during autoregressive training can also lead to unfairness, even when the SID representations are balanced. Furthermore, although fairness optimization may slightly affect recommendation accuracy on a few settings, the complete framework maintains comparable Recall and NDCG in most cases.

\begin{table*}[!t]
\centering
\small
\setlength{\tabcolsep}{0.95mm}

\begin{tabular}{clccccccccccccccc}
\toprule

& & \multicolumn{5}{c}{Beauty}
& \multicolumn{5}{c}{Industrial}
& \multicolumn{5}{c}{Software} \\

\cmidrule(lr){3-7}
\cmidrule(lr){8-12}
\cmidrule(lr){13-17}
& Method
& R@5 & R@10 & N@5 & N@10 & G@10
& R@5 & R@10 & N@5 & N@10 & G@10
& R@5 & R@10 & N@5 & N@10 & G@10 \\

\midrule

\multirow{4}{*}{\rotatebox[origin=c]{90}{Llama3.1}}
& Raw
& \textbf{0.3116} & \textbf{0.3234} & \textbf{0.2855} & \textbf{0.2894} & 0.7174
& 0.0904 & \underline{0.0970} & \underline{0.0839} & \underline{0.0861} & 0.6059
& \underline{0.2139} & 0.2386 & 0.1834 & 0.1913 & 0.8610 \\

& w/o BSQ
& 0.3048 & 0.3153 & \underline{0.2817} & \underline{0.2852} & 0.6937
& 0.0907 & 0.0967 & \textbf{0.0847} & \textbf{0.0866} & 0.5630
& 0.2030 & 0.2287 & 0.1760 & 0.1845 & 0.8454 \\

& w/o HFC
& 0.3048 & 0.3145 & 0.2799 & 0.2831 & \underline{0.5407}
& \underline{0.0913} & \textbf{0.0973} & 0.0828 & 0.0847 & \underline{0.4211}
& 0.2112 & \textbf{0.2419} & \underline{0.1840} & \underline{0.1940} & \underline{0.7089} \\

& Full
& \underline{0.3066} & \underline{0.3164} & 0.2801 & 0.2832 & \textbf{0.4976}
& \textbf{0.0918} & 0.0967 & 0.0835 & 0.0851 & \textbf{0.3439}
& \textbf{0.2145} & \underline{0.2414} & \textbf{0.1866} & \textbf{0.1952} & \textbf{0.6754} \\

\midrule

\multirow{4}{*}{\rotatebox[origin=c]{90}{Qwen3}}
& Raw
& 0.3158 & 0.3353 & 0.2895 & 0.2957 & 0.7368
& \textbf{0.0908} & \textbf{0.0989} & \underline{0.0839} & \underline{0.0865} & 0.6559
& 0.2090 & 0.2414 & \underline{0.1840} & 0.1944 & 0.8646 \\

& w/o BSQ
& 0.3164 & 0.3353 & 0.2877 & 0.2939 & 0.7108
& 0.0906 & \underline{0.0979} & \textbf{0.0842} & \textbf{0.0866} & 0.6054
& \underline{0.2123} & 0.2534 & 0.1815 & 0.1946 & 0.8695 \\

& w/o HFC
& \underline{0.3284} & \underline{0.3481} & \underline{0.2937} & \underline{0.3002} & \underline{0.5563}
& \underline{0.0907} & 0.0973 & 0.0811 & 0.0832 & \underline{0.4858}
& 0.2095 & \underline{0.2567} & 0.1815 & \underline{0.1967} & \underline{0.7350} \\

& Full
& \textbf{0.3329} & \textbf{0.3521} & \textbf{0.2960} & \textbf{0.3023} & \textbf{0.5128}
& 0.0891 & 0.0971 & 0.0807 & 0.0833 & \textbf{0.4203}
& \textbf{0.2189} & \textbf{0.2639} & \textbf{0.1872} & \textbf{0.2019} & \textbf{0.7127} \\

\bottomrule
\end{tabular}
\caption{Ablation study of FSGR. Removing either BSQ (w/o BSQ) or HFC (w/o HFC) degrades recommendation fairness, while the complete framework (Full) achieves the best overall trade-off between accuracy and fairness.}
\label{tab:ablation}
\end{table*}

\begin{figure}[!ht]
  \centering
  \includegraphics[width=0.95\linewidth]{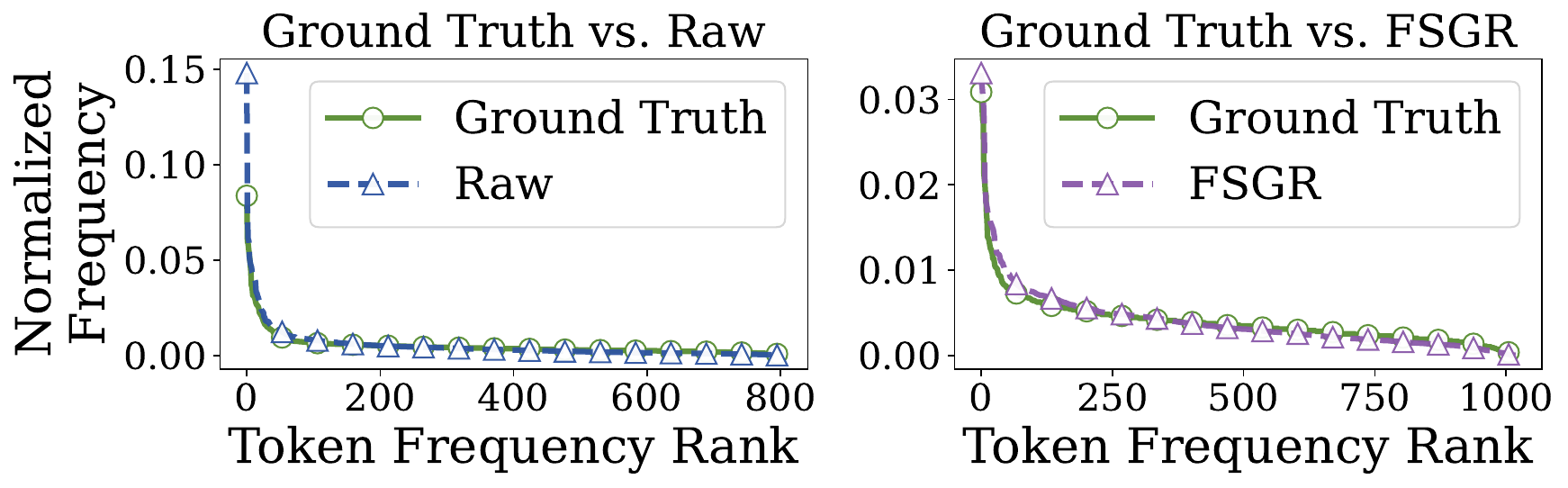}
    \caption{Comparison of SID token frequency distributions between model predictions and ground-truth on Industrial using Qwen3. FSGR produces a smoother distribution that better matches the ground-truth, mitigating high-frequency bias observed in raw RQ-VAE and LLM.}
  \label{fig:Token_Frequency_Distribution}
\end{figure}

\subsection{Token Frequency Distribution Analysis}

Figure~\ref{fig:Token_Frequency_Distribution} compares frequency distributions of SID tokens in the generated recommendation results with the ground-truth distribution on Qwen3. The raw RQ-VAE and LLM overestimates the most frequent SID tokens, producing a much steeper distribution than the ground truth, which reflects the token frequency bias identified in this work. In contrast, FSGR generates a distribution closely matches the target distribution, substantially reducing the over-prediction of high-frequency tokens while increasing the relative exposure of less frequent tokens. This demonstrates that the proposed BSQ and HFC alleviate token frequency bias throughout both SID construction and recommendation generation.

\subsection{Analysis of Codebook Utilization}

\begin{figure}[!h]
  \centering
  \includegraphics[width=0.95\linewidth]{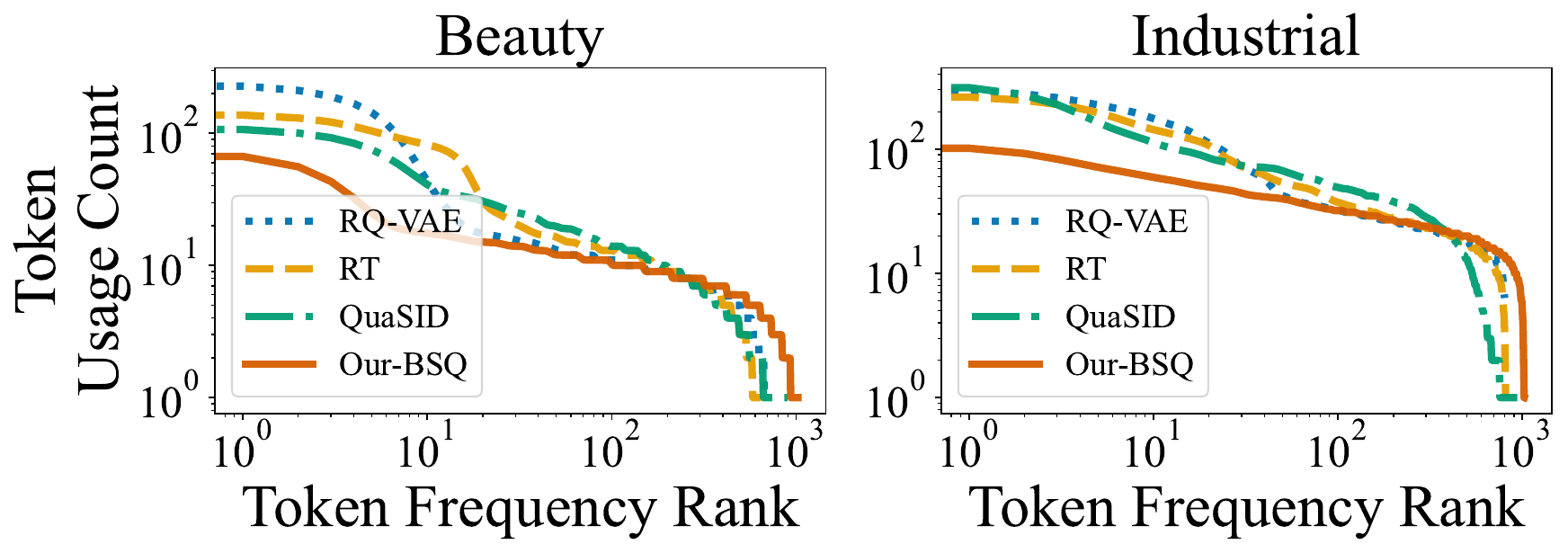}
    \caption{Token frequency distributions of different Semantic ID construction methods on two datasets. 
    Tokens are sorted by usage frequency in descending order.
    A flatter distribution indicates more balanced codebook utilization.}
  \label{fig:SID_Token_Freq}
\end{figure}

To evaluate whether BSQ balances SID allocation, we analyzed codebook utilization from both statistical and distributional perspectives. Table~\ref{tab:sid} reports the codebook Coverage and Gini coefficient, while Figure~\ref{fig:SID_Token_Freq} illustrates the token frequency distributions sorted by usage frequency. As shown in Table~\ref{tab:sid}, the proposed method achieves the highest Coverage and the lowest Gini on all datasets. In particular, the Coverage approaches $100\%$ on Industrial dataset, indicating that nearly all codewords participate in quantization. Meanwhile, the lower Gini scores demonstrate that token usage is distributed much more evenly across the codebook than that of other methods. Figure~\ref{fig:SID_Token_Freq} further provides a distributional view of codebook utilization. Compared with the baselines, the proposed method exhibits a flatter token frequency curve, indicating that the dominance of a small number of high-frequency codewords is alleviated while previously under-utilized codewords are activated more frequently. These results demonstrate that BSQ encourages a more balanced semantic representation space, providing a stronger foundation for mitigating token frequency bias in downstream generative recommendation.

\begin{table}[!h]
\small

\centering
\small
\setlength{\tabcolsep}{1.4mm}
\begin{tabular}{llcccc}
\toprule
Dataset & Metric & RQ-VAE & RT & QuaSID & Our-BSQ \\
\midrule

\multirow{2}{*}{Beauty}
& Coverage & 0.6748 & 0.6309 & \underline{0.8779} & \textbf{0.9990} \\
& Gini     & 0.6303 & 0.6864 & \underline{0.6119} & \textbf{0.3479} \\
\midrule

\multirow{2}{*}{Industrial}
& Coverage & 0.7832 & 0.7930 & \underline{0.9180} & \textbf{1.0000} \\
& Gini     & \underline{0.4904} & 0.5107 & 0.6064 & \textbf{0.2590} \\
\midrule

\multirow{2}{*}{Software}
& Coverage & 0.4639 & 0.4707 & \underline{0.8623} & \textbf{0.9824} \\
& Gini     & 0.7954 & 0.7972 & \underline{0.6199} & \textbf{0.5138} \\
\bottomrule
\end{tabular}
\caption{Comparison of codebook utilization across different Semantic ID construction methods. Higher Coverage and lower Gini indicate more balanced codebook utilization.}
\label{tab:sid}
\end{table}

\subsection{Analysis of Layer-wise Temperature Assignment}

\begin{table}[!t]
\centering

\small
\begin{tabular}{llccc}
\toprule
Dataset & Metric & HFC & Reverse $\tau_l$ & Fixed $\tau_l$ \\
\midrule

\multirow{5}{*}{Beauty}
& R@5      & \textbf{0.3329} & 0.3258 & \underline{0.3316} \\
& R@10     & \textbf{0.3521} & 0.3452 & \underline{0.3471} \\
& N@5   & \underline{0.2960} & 0.2921 & \textbf{0.2964} \\
& N@10  & \textbf{0.3023} & 0.2986 & \underline{0.3014} \\
& G@10  & 0.5128 & \textbf{0.4921} & \underline{0.5094} \\
\midrule

\multirow{5}{*}{Industrial}
& R@5      & \underline{0.0891} & 0.0869 & \textbf{0.0899} \\
& R@10     & \textbf{0.0971} & 0.0941 & \underline{0.0963} \\
& N@5   & \underline{0.0807} & 0.0796 & \textbf{0.0810} \\
& N@10  & \textbf{0.0833} & 0.0820 & \underline{0.0830} \\
& G@10  & 0.4203 & \textbf{0.3228} & \underline{0.3632} \\
\midrule

\multirow{5}{*}{Software}
& R@5      & \textbf{0.2189} & \underline{0.2068} & 0.2063 \\
& R@10     & \textbf{0.2639} & \underline{0.2479}& 0.2436 \\
& N@5   & \textbf{0.1872} & 0.1815 & \underline{0.1818} \\
& N@10  & \textbf{0.2019} & \underline{0.1946} & 0.1937 \\
& G@10  & \textbf{0.7127} & \underline{0.7218} & 0.7342 \\
\bottomrule
\end{tabular}
\caption{Comparison of different layer-wise temperature assignment strategies for HFC on Qwen3. Reverse $\tau_l$ assigns larger calibration temperatures to lower SID layers, Fixed $\tau_l$ uses the same temperature for all layers.}
\label{tab:hfc}
\end{table} 

To validate the effectiveness of the proposed layer-wise temperature assignment ($\tau_l=l/L$), we conducted experiments on Qwen3,  comparing the HFC with two alternative strategies: \textbf{Reverse $\tau_l$}, which applies larger calibration temperatures to lower SID layers and smaller ones to higher layers ($\tau_l=(L-l+1)/L$), and \textbf{Fixed $\tau_l$}, which assigns the same calibration strength to all SID layers ($\tau_l=(\Sigma_{1}^{L}l/L)/L$). The results are reported in Table~\ref{tab:hfc}. Overall, the proposed HFC achieves the best overall trade-off between recommendation accuracy and fairness across all three datasets. Although reverse $\tau_l$ further reduces the Gini coefficient on Beauty and Industrial, it degrades Recall and NDCG, indicating that applying strong calibration to lower SID layers disrupts coarse semantic representations. In contrast, the proposed layer-wise schedule applies weaker calibration to lower layers and progressively strengthens it for deeper layers, effectively mitigating frequency bias while preserving semantic alignment. Compared with using fixed $\tau_l$, HFC achieves better overall performance, demonstrating that different SID layers require different calibration strengths according to their hierarchical semantic roles. These results validate the effectiveness of the proposed layer-aware temperature assignment.

\section{Conclusion}
In this paper, we investigate the token frequency bias problem in Semantic ID (SID)-based generative recommendation and propose FSGR, a fairness optimization framework that jointly improves SID construction and recommendation training. Specifically, FSGR combines Balanced Semantic Quantization for SID generation with Two-Stage Recommendation Training to mitigate frequency bias during recommendation training. Experiments on three datasets and three backbones demonstrate that FSGR improves SID token fairness while maintaining competitive recommendation accuracy on most settings. In future work, we plan to extend FSGR to more advanced generative recommendation architectures and explore adaptive frequency-aware optimization strategies for better fairness–accuracy trade-offs.

\bibliography{aaai2027}


\end{document}